\documentclass[aip, reprint]{revtex4-1}
\usepackage{graphicx}
\usepackage{floatrow}
\usepackage{dcolumn}
\usepackage{bm}
\usepackage{amsmath}

\begin{document}

\title{Anisotropy, band-to-band transitions, phonon modes, and oxidation properties of cobalt-oxide core-shell slanted columnar thin films} 

\author{Alyssa Mock}
\email[Electronic mail: ]{amock@huskers.unl.edu}
\affiliation{Department of Electrical Engineering and Center for Nanohybrid Functional Materials, University of Nebraska-Lincoln, Lincoln, NE 68588, USA}
\author{Rafa\l{} Korlacki}
\affiliation{Department of Electrical Engineering and Center for Nanohybrid Functional Materials, University of Nebraska-Lincoln, Lincoln, NE 68588, USA}
\author{Chad Briley}
\affiliation{Department of Electrical Engineering and Center for Nanohybrid Functional Materials, University of Nebraska-Lincoln, Lincoln, NE 68588, USA}
\author{Derek Sekora}
\affiliation{Department of Electrical Engineering and Center for Nanohybrid Functional Materials, University of Nebraska-Lincoln, Lincoln, NE 68588, USA}
\author{Tino Hofmann}
\affiliation{Department of Electrical Engineering and Center for Nanohybrid Functional Materials, University of Nebraska-Lincoln, Lincoln, NE 68588, USA}
\affiliation{Department of Physics, Chemistry, and Biology, Link$\ddot{o}$ping University, Link$\ddot{o}$ping, Sweden}
\author{Peter Wilson}
\affiliation{Department of Chemistry and Center for Nanohybrid Functional Materials, University of Nebraska-Lincoln, Lincoln, NE 68588, USA}
\author{Alexander Sinitskii}
\affiliation{Department of Chemistry and Center for Nanohybrid Functional Materials, University of Nebraska-Lincoln, Lincoln, NE 68588, USA}
\author{Eva Schubert}
\affiliation{Department of Electrical Engineering and Center for Nanohybrid Functional Materials, University of Nebraska-Lincoln, Lincoln, NE 68588, USA}
\author{Mathias Schubert}
\affiliation{Department of Electrical Engineering and Center for Nanohybrid Functional Materials, University of Nebraska-Lincoln, Lincoln, NE 68588, USA}

\date{\today}

\begin{abstract}
Highly-ordered and spatially-coherent cobalt slanted columnar thin films were deposited by glancing angle deposition onto silicon substrates, and subsequently oxidized by annealing at 475~$^\circ$C. Scanning electron microscopy, Raman scattering, generalized ellipsometry, and density functional theory investigations reveal shape-invariant transformation of the slanted nanocolumns from metallic to transparent metal-oxide core-shell structures with properties characteristic of spinel cobalt oxide. We find passivation of Co-SCTFs yielding Co-Al$_2$O$_3$ core-shell structures produced by conformal deposition of a few nanometers of alumina using atomic layer deposition fully prevents cobalt oxidation in ambient and from annealing up to 475~$^\circ$C.
\end{abstract}

\maketitle

\begin{figure}[hbt]
    \includegraphics[width=0.4\linewidth]{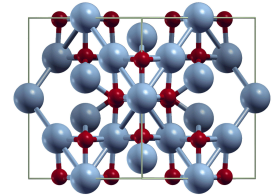}
    \caption{A unit cell of Co$_3$O$_4$ (spinel structure) viewed along the [110] direction. (Ref \onlinecite{Kokalj2003})}
    \label{unitcell}
\end{figure}

Transition metal oxides constitute an interesting class of materials in solid state physics, with numerous attributes that span dielectric, semiconducting, ferromagnetic, and ferroelectric properties. Much research effort has focused on this vast and still widely uncharted field of oxide compounds. In addition to their unique and useful bulk properties, in recent years strong interest has developed in shaping these materials into topographies with nanoscale dimensions. One example is core-shell nanostructures whose optical, electrical, and magnetic properties are attractive for novel device architectures in applications such as solar cells and sensors.\cite{Zhao2015, Chen2006, Jiang2012, Su2011, Yang2011} Nanostructures and core-shell structures that incorporate the transition metal semiconductor cobalt oxide have been shown to be effective as anodes with largely increased surface area in lithium ion batteries, biosensors, and in electrochemical catalysis.\cite{Li2005, Wang2009, Salimi2009, Ibupoto2014, Jiao2010, Benitez2008} This is due to the wide tunability of electrical and magnetic properties as a function of temperature and stoichiometry. Oxidation of cobalt may result in different forms depending on growth conditions. Rock salt structure cobaltous oxide (CoO) is known to be present at high temperatures from decomposition of cobaltic oxide (Co$_{2}$O$_{3}$) or spinel structure mixed-valence compound Co$^{II}$Co$^{III}$$_{2}$O$_{4}$ (Co$_{3}$O$_{4}$) while at lower temperatures increased oxygen absorption produces Co$_{3}$O$_{4}$ (Fig.~\ref{unitcell}).\cite{Gulbransen1951, HutchingsCobaltOxide2006, Metz2009}

\begin{figure*}[hbt]
    \includegraphics[width=0.9\linewidth]{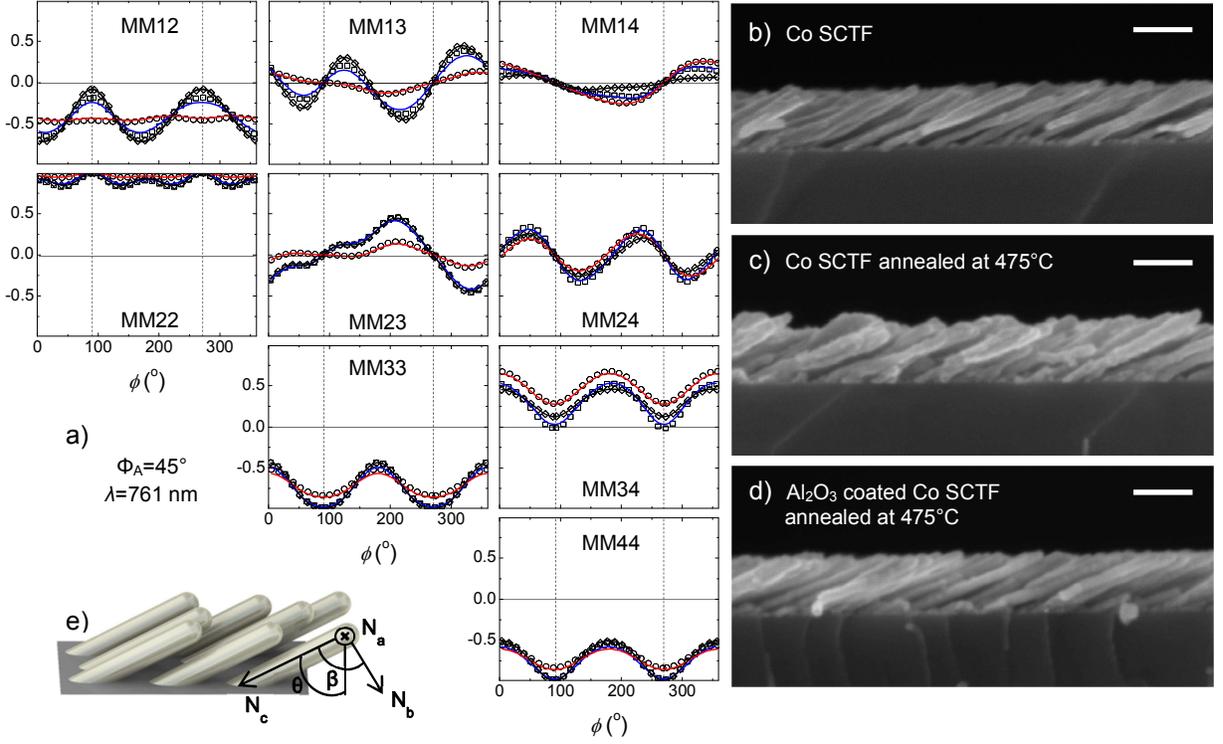}
    \caption{(a) Exemplary experimental (scattered lines) and HBLA calculated (solid lines) MMGE data at an angle of incidence $\phi_A=45^\circ$ and wavelength $\lambda=761$~nm of as grown Co SCTF (square/blue), Co SCTF becoming Co$_3$O$_4$ after annealing (circle/red), and Al$_2$O$_3$ coated Co SCTF after annealing (diamond/black) are presented versus in-plane sample azimuth, $\phi$. The zero-crossing points in the off-block diagonal elements (MM13, MM14, MM23, MM24) indicate when the slanted columns are parallel to the plane of incidence. Cross-sectional SEM images of cobalt SCTF (b) as grown, (c) annealed becoming Co$_3$O$_4$, and (d) annealed after Al$_2$O$_3$ passivation. All scale bars represent 100~nm. (e) A graphic detailing the coordinate axes used in this paper to describe an SCTF.}
    \label{SEMpic}
\end{figure*}

In this paper we report on fabrication and characterization of highly regular and coherently arranged few-nanometer-sized columnar core-shell structures consisting of metal cores and metal-oxide shells (Figs.~\ref{SEMpic}(b)-\ref{SEMpic}(d)). In particular, we investigate Co nanocolumns and how their physical properties change upon oxidation. In order to avoid possible ambiguity we investigate and report optical, vibrational, electronic, and oxidation properties. We find that cobalt nanocolumns can be almost completely transformed into Co$_{3}$O$_{4}$ while fully retaining their shape. Thereby, we have produced highly anisotropic spatially coherent cobalt oxide nanocolumnar thin films and are able to discuss their physical properties which may be useful for potential optical and/or electrochemical device architectures. We further find that passivation of cobalt nanocolumns by an ultra-thin conformal overlayer of alumina fully inhibits oxidation of the cobalt nanocolumns at normal ambient as well as during our annealing procedure similar to our previous results with conformal coating of a few-layers of graphene.\cite{Wilson2015, Wilson2015a, Wilson2014}

Highly-ordered, spatially-coherent slanted columnar thin films (SCTF) can be fabricated by glancing angle deposition (GLAD) which utilizes particle flux at oblique incidence angles.\cite{Hawkeye2007} These films exhibit extreme birefringence and dichroism properties that can be tailored by choice of deposition parameters.\cite{Hawkeye2014} An ultra-thin and conformally over-grown passivation layer via atomic layer deposition (ALD) can be employed to adjust physical and chemical surface properties of SCTFs.\cite{SchmidtPassivation2012, Ylivaara2014, Knez2007} ALD is a chemical vapor deposition process which provides precise, self-limiting, monolayer growth of materials by cycling a combination of precursor, ozone, or plasma.

\begin{table}[hbt]
\begin{tabular}{l | c | c | c | c}
\hline \hline
\multicolumn{1}{c|}{} & \multicolumn{1}{c|}{} & \multicolumn{1}{c|}{Co} & \multicolumn{1}{c|}{Co-Co$_3$O$_4$} & \multicolumn{1}{c}{Co-Al$_2$O$_3$} \\
\ & Parameter & As Grown & Annealed & Annealed \\
\hline
\ & $t$(nm) & 99(9) & 97(8) & 98(4) \\
SEM & $\theta(^\circ$) & 64(4)  & 63(4) & 64(3) \\
\ & $d$(nm) & 17(2) & 26(6) & 22(4) \\
\hline
\ & $t$ (nm) & 83.4(1) & 82.0(1) & 97.2(1) \\
\ & $\theta$ ($^\circ$)& 61.4(1) & 61.7(1) & 64.2(1) \\
MMGE & $\beta$ ($^\circ$) & 79.6(1) & 83.2(1) & 77.5(1) \\
\ & Co \% & 24.9(1) & 2.3(1) & 27.6(1) \\
\ & Co$_3$o$_4$ \% & - & 42.6(1) & - \\
\ & Al$_2$O$_3$ \% & - & - & 16.8(2) \\
\hline \hline
\end{tabular}
\caption{SCTF parameters obtained by SEM and MMGE. Numbers in parentheses are uncertainty limits corresponding to the confidence level of at least 90\% with respect to the last reported digits.}
\label{Data}
\end{table}

Mueller matrix generalized ellipsometry (MMGE) determines the anisotropic optical properties of SCTFs by measurement and model analysis of the 4x4 Mueller matrix \textbf{M}, where \textbf{M} describes the change in polarization of light after interaction with a sample (Fig.~\ref{SEMpic}(a)).\cite{Schmidtthinsolidfilms2013, Fujiwara} An appropriate physical model then allows for accurate description of the optical properties of the film. The dielectric function of a uniform flat film may be determined using a homogeneous isotropic layer approach (HILA). For anisotropic films such as SCTFs, the dielectric functions along all three major axes \textbf{N$_{\textnormal{a}}$}, \textbf{N$_{\textnormal{b}}$}, and \textbf{N$_{\textnormal{c}}$} (Fig.~\ref{SEMpic}(e)) may be obtained using a homogeneous biaxial layer approach (HBLA). This method assumes that the material is homogeneously distributed along each distinct axis and does not allow for determination of material constituents.\cite{SchmidtBruggeman2013} For columnar core-shell structures, the anisotropic Bruggeman effective medium approximation (AB-EMA) can be used to model the dielectric function of all three major axes from several "bulk" material models in addition to the constituent fractions according to:
\begin{equation}
\sum_{n=1}^{m} f_n \frac{\varepsilon_{n}-\varepsilon_{\textrm{eff},j}}{\varepsilon_{\textrm{eff},j}+L_j^D(\varepsilon_n - \varepsilon_{\textrm{eff},j})} = 0, \label{Brug}
\end{equation}
with the requirements that $\sum f_n = 1$ and $\sum L_j^D =1$, where $L_j^D$ are the depolarization factors along the major axes, \textbf{N$_{\textnormal{a}}$}, \textbf{N$_{\textnormal{b}}$}, and \textbf{N$_{\textnormal{c}}$}.\cite{SchmidtBruggeman2013, Granqvist1977} For the AB-EMA model we assumed three different constituents: void, cobalt, and oxide, either the well documented Al$_{2}$O$_{3}$ deposited by ALD or the as yet to be described cobalt oxide.\cite{SchmidtBruggeman2013} The optical constants of the cobalt SCTF constituent were determined from the as-deposited sample immediately after deposition, where there was assumed to be no oxide material present. These were then kept constant during the search for the unknown oxide optical constants of the core-shell structures. Furthermore, we previously reported that SCTFs exhibit monoclinic properties,\cite{SchmidtPassivation2012} which can be accounted for by defining an angle $\beta$ between axes \textbf{N$_{\textnormal{b}}$}, and \textbf{N$_{\textnormal{c}}$} (Fig.~\ref{SEMpic}(b)). During best-match model calculations model parameters are then varied in order to calculate data that matches the experimental data as accurately as possible. Values of the complex dielectric function of an unknown material can be found by a wavelength-by-wavelength regression analysis. Electronic band-to-band transitions cause critical point (CP) features in the dielectric function spectra. The imaginary contribution, $\varepsilon_2$, can be modeled conveniently using Gaussian lineshapes with best-match parameters amplitude, A, center energy, E$_n$, and broadening, B$_r$, and the real contribution, $\varepsilon_1$, is obtained from Kramers-Kronig transformation.\cite{Meneses2006}
\begin{eqnarray}
\varepsilon_2(E)& = &A(e^{-(\frac{E-E_n}{\sigma})^2}-e^{-(\frac{E+E_n}{\sigma})^2}), \label{e2} \\
\sigma& = &\frac{B_r}{2\sqrt{ln(2)}},  \nonumber \\
\varepsilon_1(E)& = &\frac{2}{\pi}P\int_0^\infty \frac{\xi\varepsilon_2(\xi)}{\xi^2-E^2}d\xi.
\end{eqnarray}
\indent Raman spectroscopy was used to differentiate between the cobalt oxide phases, however, it has been shown that at high excitation laser energy in normal ambient, CoO can further oxidize leading to Co$_{3}$O$_{4}$. Care must be taken to use a low enough intensity so as to not change the material investigated.\cite{Gallant2006} Near the Brillouin zone center (k=0) of Co$_3$O$_4$ (space group O$_h$$^7$)\cite{Rousseau1981} normal modes are given from:
\begin{equation}
\Gamma = A_{1g} + E_g + 3F_{2g} + 5F_{1u} + 2A_{2u} + 2E_u + 2F_{2u},
\label{Modes}
\end{equation}
where the A$_1$$_g$, E$_g$, and the triple degenerate 3F$_2$$_g$ are the only Raman active modes. \cite{Hadjiev1988}

\begin{figure}[hbt]
    \includegraphics[width=\linewidth]{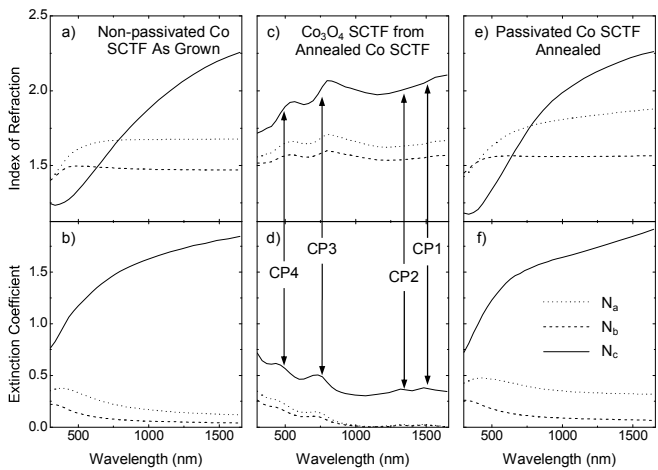}
    \caption{Optical constants along major axes \textbf{N$_{\textnormal{a}}$}, \textbf{N$_{\textnormal{b}}$}, and \textbf{N$_{\textnormal{c}}$} of non-passivated (a)-(d) and passivated (e)-(f) cobalt SCTFs. Critical point transition energies, identified from the subsequent AB-EMA analysis, are indicated by vertical lines.}
    \label{effoptconst}
\end{figure}

Density functional theory (DFT) calculations of the Raman spectrum of Co$_3$O$_4$ were performed using plane-wave code \textit{Quantum Espresso} (QE).\cite{qe} Local density approximation functions by Perdew and Zunger\cite{PZ} (PZ) and norm-conserving pseudopotentials from the QE library were used, with the cutoff for the electronic wave-function set at 80 Ry, and a $4 \times 4 \times 4$ Monkhorst-Pack grid shifted by half of the simulation cell for the Brillouin-Zone integrations.\cite{mp} The primitive unit cell was first relaxed in order to reduce forces on the ions. The system is considered to be at equilibrium when the forces on the ions were approximately 1.0 $\times$ 10$^{-5}$ Ry/bohr ($\sim$ 0.0003 eV/\AA). The phonon frequencies and Raman activities were computed at the $\Gamma$ point for the relaxed structure using density-functional perturbation-theory for phonons\cite{DFPT} and the second-order response for Raman activities.\cite{lazzeri03} 


\begin{figure}[hbt]{\includegraphics[width=\linewidth]{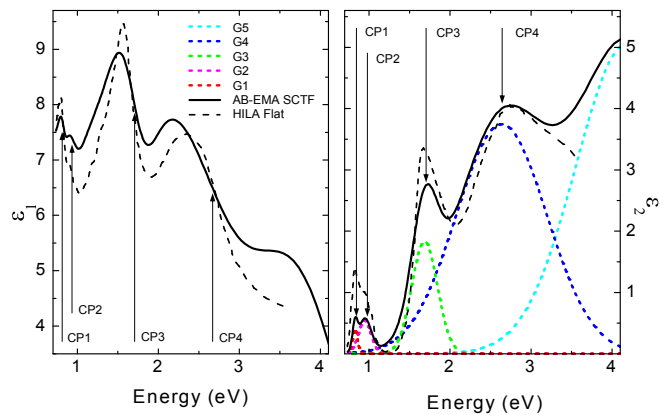}}
    \caption{AB-EMA model dielectric function along axis \textbf{N$_{\textnormal{c}}$} of non-passivated annealed Co-SCTF along with Gaussian curves from CP analysis. A HILA model dielectric function of Co$_3$O$_4$ obtained from data taken on a soda-lime float glass after a 12 second pyrolytic spray deposition by Athey \textit{et. al} (Ref. \onlinecite{AtheyOpticalPropertiesCo3O41996}) are shown for comparison.}
    \label{oxide}
\end{figure}

\begin{table*}[hbt]
\setlength{\tabcolsep}{12pt}
{\begin{tabular}{l | c | c | c | c | c }
\hline \hline
&\multicolumn{5}{c}{Critical Point Transitions}\\
\multicolumn{1}{c|}{Parameter} & \multicolumn{1}{c}{CP1} & \multicolumn{1}{c}{CP2} & \multicolumn{1}{c}{CP3} & \multicolumn{1}{c}{CP4} & \multicolumn{1}{c}{CP5} \\
\hline
E$_n$(eV) & 0.827(1) & 0.947(1) & 1.686(1) & 2.637(1) & 4.187(1) \footnote{Transition outside of presented range with limited sensitivity to critical point parameters.} \\
B$_r$(eV)& 0.072(1) & 0.181(1) & 0.373(1) & 1.313(1) & 1.491(1) \\
A(eV) & 0.405(1) & 0.541(1) & 1.834(1) & 3.745(1) & 5.058(1) \\
E(eV)\footnote{Range of parameter values presented for spray pyrolysis depositions.} [Ref. \onlinecite{AtheyOpticalPropertiesCo3O41996}]   & 0.83 - 0.85 & 0.93 - 0.98 & 1.70 - 1.72 & 2.82 - 3.10 & - \\
E(eV) [Ref. \onlinecite{Miedzinska1987}]  & 0.8 & 0.9 & 1.6 & 2.8 & - \\
E(eV) [Ref. \onlinecite{Martens1985}] & 0.8 & - & 1.6 & 2.65 & 4.4 \\
E(eV) [Ref. \onlinecite{Cook1986}]  & 0.82 & 0.93 & 1.7 & 2.8 & - \\
E(eV) [Ref. \onlinecite{Kormondy2014}]  & 0.9\footnote{Double peak was suggested around 0.9~eV.} & 0.9 & 1.65 & 2.6 & 5 \\
\hline \hline
\end{tabular}}
\caption{Critical point transition parameters of Co$_{3}$O$_{4}$ obtained in this work from MMGE data analysis of oxidized cobalt SCTF in comparison with selected data reported in literature.}
\label{Gausparm}
\end{table*}

Cobalt slanted columnar thin films (SCTF) were deposited on Si(100) substrates with an intrinsic 1.7 nm layer of native oxide using GLAD as described previously,\cite{SchmidtBruggeman2013} and at room temperature under ultra-high vacuum with an oblique angle of incidence of 85$^\circ$. Half of the samples were passivated by depositing a conformal layer of Al$_{2}$O$_{3}$ using documented ALD techniques \cite{Ylivaara2014, SchmidtPassivation2012} by subsequent cycling of trimethylaluminum (TMA) and 18.3~M$\Omega$ deionized water (Fiji 200 ALD Reactor, Cambridge NanoTech). The samples were held at 150~$^\circ$C temperature under vacuum, and 55 cycles were deposited with a rate of approximately 0.9 \AA/cycle.\cite{SchmidtPassivation2012} All samples were characterized immediately after deposition using SEM and MMGE to determine initial thickness, column diameter, and optical properties. MMGE data were measured in the spectral range of 300 to 1650 nm at four angles of incidence, $\phi_{A}=45^\circ, 55^\circ, 65^\circ,$ and $75^\circ$, and over a full azimuthal rotation of the sample by six degree increments (RC2, J.~A.~Woollam Co, Inc.). All samples were then placed flat on a sample mount plate under high vacuum and heated gradually to 475~$^\circ$C by increments of 25~$^\circ$C every 30 minutes, held at 475~$^\circ$C for 2 hours, then suddenly exposed to normal ambient, and then allowed to cool to room temperature. SEM and MMGE investigations were then repeated, and Raman spectroscopy was performed with a 532 nm excitation laser (DXR Raman Microscope, ThermoScientific).

Cross-sectional SEM images are presented in Figs.~\ref{SEMpic}(b)-\ref{SEMpic}(d). Approximations for the overall thickness of the SCTF (t), as well as the slanting angle ($\theta$) and diameter of the posts (d) are summarized in Tab.~\ref{Data}. Upon annealing the column diameter of the non-passivated SCTF increases approximately twofold due to oxide growth on all exposed surfaces, while the column diameter of the passivated sample is approximately 5~nm larger than that of the as-grown columns due to the alumina overgrowth. Otherwise no significant structural changes occur upon annealing or passivation.

\begin{figure}[hbt]{\includegraphics[width=0.9\linewidth]{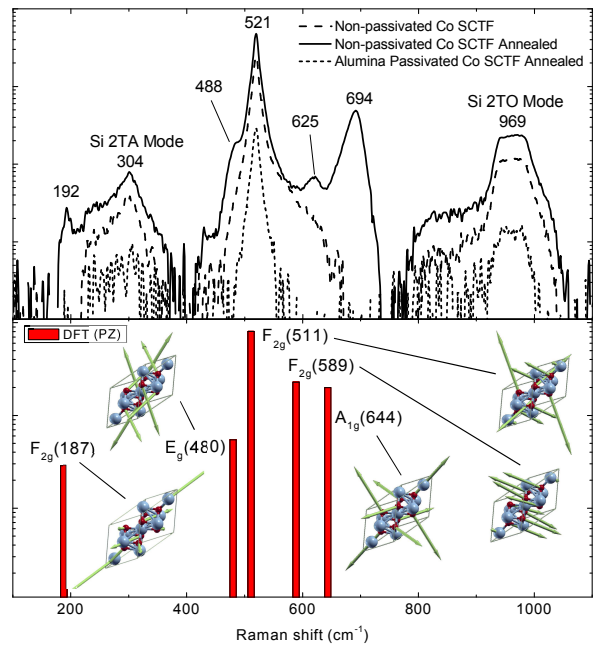}}
    \caption{Raman spectroscopy experimental results for non-passivated cobalt SCTF after each annealing and exposure step in comparison with DFT simulation results for cobalt oxide, Co$_{3}$O$_{4}$, with examples of mode displacement vectors for A$_1$$_g$, E$_g$, and triple-degenerate 3F$_2$$_g$ modes.}
    \label{Raman}
\end{figure}

Fig.~\ref{SEMpic}(a) compares MMGE data versus sample rotation with HBLA best-match model calculated data for as grown, non-passivated annealed, and passivated annealed SCTFs. It can be seen that the data from the passivated annealed SCTF very closely resemble those of the as grown SCTF, signifying that both SCTFs remained structurally and optically nearly identical. While data from the non-passivated annealed SCTF reveal some deviation from the former two, major anisotropic properties of the SCTFs remained similar upon oxidation. This is seen by observing where the zero points in the off-block diagonal elements in the Mueller Matrix data in Fig.~\ref{SEMpic}(a) occur, in which the slanted columns are aligned within the plane of incidence causing pseudo-isotropic points.\cite{SchmidtBruggeman2013} For all SCTFs, pseudo-isotropic points occur at the same azimuthal orientations. Best-match AB-EMA model parameters are summarized in Tab.~\ref{Data}. SCTF volume fractions for cobalt and oxide change with temperature and lead to an increase in diameter of the columns. Before annealing, the SCTF is assumed to have no oxide formation yielding a void fraction of about 75\%. It is observed that after annealing the original metal SCTF is almost fully transformed into metal oxide forming a shell around a reducing core with remaining diameter of approximately 2~nm still above our best-match model parameter uncertainty range (Tab.~\ref{Data}). For Co$_3$O$_4$ the density is less than that of pure cobalt, thus its formation results in an increase in column diameter and a reduction of the void fraction to approximately 55\%, from which we suggest a bidirectional oxide growth. Fig.~\ref{effoptconst} depicts the anisotropic optical constants (refractive index and extinction coefficient). The optical constants along \textbf{N$_{\textnormal{a}}$}, \textbf{N$_{\textnormal{b}}$}, and \textbf{N$_{\textnormal{c}}$} change between the as-grown and the non-passivated annealed SCTF, while those of the passivated annealed SCTF remain nearly identical to those of the as-grown SCTF. The oxidized cobalt SCTF exhibits strong changes in birefringence and dichroism properties, transforming from highly absorbing (metal-like) to highly transparent (dielectric-like). For polarization along axis \textbf{N$_{\textnormal{c}}$} the optical constants are typically found nearly identical to the bulk optical constants of the column material, here bulk cobalt. For annealed non-passivated SCTF, we thus indicate the transformation to Co$_3$O$_4$ because the optical constants for axis \textbf{N$_{\textnormal{c}}$} are nearly identical to bulk Co$_3$O$_4$. The AB-EMA optical constants for axis \textbf{N$_{\textnormal{c}}$} are compared to those given by Athey \textit{et. al},\cite{AtheyOpticalPropertiesCo3O41996} which were obtained using the HILA model for data from a Co$_{3}$O$_{4}$ thin film deposited using 12 second spray pyrolysis onto soda-lime float glass, included in Fig.~\ref{oxide}. These results show close agreement on magnitudes as well as absorption band locations yielding further confirmation that cobalt oxide is in the form of Co$_{3}$O$_{4}$. Our CP parameter analysis results yield optical constants for axis \textbf{N$_{\textnormal{c}}$} which are compared to those reported previously in Tab.~\ref{Gausparm}.\cite{AtheyOpticalPropertiesCo3O41996, Miedzinska1987, Martens1985, Cook1986, Kormondy2014} Raman spectra of the non-passivated annealed SCTF with DFT calculated Co$_3$O$_4$ Raman intensities are shown in Fig.~\ref{Raman}. We identify peaks that correspond to modes F$^1$$_2$$_g$, E$_g$, F$^2$$_2$$_g$, F$^3$$_2$$_g$, and A$_1$$_g$ at 192~cm$^{-1}$, 488~cm$^{-1}$, 521~cm$^{-1}$, 625~cm$^{-1}$, and 694~cm$^{-1}$, respectively. We note that the peak at 521~cm$^{-1}$ is caused by the silicon substrate and the Co$_3$O$_4$ peak at 488~cm$^{-1}$ is subsumed as a shoulder. We also make note that the Co$_3$O$_4$ peak at 521~cm$^{-1}$ is not visible as it appears at the same frequency as the strong silicon peak. Surface-enhanced scattering effects cause the secondary Si modes, 2TA at 304~cm$^{-1}$ and 2TO at 969~cm$^{-1}$, which pronounce upon oxidation due to the increased SCTF transparency.\cite{Gonzalez2014} Fig.~\ref{Raman} also shows examples of mode displacement vectors for A$_1$$_g$, E$_g$, and triple-degenerate 3F$_2$$_g$ modes. We find very good agreement between experimental and DFT predicted Raman intensities as well as with literature for Co$_3$O$_4$.\cite{Yu2005, Hadjiev1988} We therefore conclude that the annealing process of non-passivated Co-SCTF leads to nearly full oxidation forming Co-Co$_3$O$_4$ core-shell structures. The passivated annealed SCTF does not reveal any mode related to cobalt oxide. We conclude that alumina passivation prevents the Co-nanocolumns from oxidation regardless of thermal annealing up to 475$^{\circ}$C, while a small diameter increase is caused by the alumina passivation layer.

In summary, we investigated the changes in optical anisotropy, the emergence of band-to-band transitions and phonon modes upon the thermal oxidation of cobalt slanted columnar thin films. A thermal annealing process was performed whereby cobalt columns were exposed to conditions favorable for oxidation in as-grown state as well as after passivation using a thin alumina layer overgrown by atomic layer deposition. We observe insignificant structural and optical changes of the passivated cobalt nanocolumns, while the non-passivated annealed columns transform into cobalt-oxide core-shell nanostructures. Scanning electron microscopy, Raman scattering, generalized ellipsometry, and density functional theory investigations reveal a shape-invariant transformation of cobalt columns from metallic to transparent metal-oxide core-shell structures with properties characteristic of spinel cobalt oxide. The anisotropic optical constants derived for the as-grown and non-passivated annealed cobalt columns reveal transformation from metal-like to dielectric-like suggesting highly transparent properties of the cobalt oxide nanostructures. From structural investigations we propose that oxide grows bidirectionally, leaving only a small fraction of a cobalt core as the remaining void fraction diminishes proportionally upon oxide formation. We conclude that alumina passivation provides an efficient oxygen barrier onto cobalt nanostructures, which may find use in design of device architectures including cobalt-cobalt oxide core-shell nanostructures for applications in normal ambient.

\begin{acknowledgments}
This work was supported by the National Science Foundation (NSF) through the Center for Nanohybrid Functional Materials (EPS-1004094) and the Nebraska Materials Research Science and Engineering Center (DMR-1420645). Partial financial support from Army Research Office (W911NF-09-C-0097), NSF (CMMI 1337856, EAR 1521428), Nebraska Center for Energy Sciences Research (Grant No. 12-00-13), and J.A. Woollam Foundation is also acknowledged. DFT calculations were performed using the resources of the Holland Computing Center at the University of Nebraska-Lincoln.
\end{acknowledgments}


%

\end{document}